\documentclass
[10pt,letterpaper,superscriptaddress,nobalancelastpage,endfloats,onecolumn,preprint,aps,titlepage]{revtex4}%
\usepackage{amsmath}
\usepackage{amssymb}
\usepackage{amsfonts}
\usepackage{graphicx}
\usepackage{hyperref}%
\providecommand{\U}[1]{\protect\rule{.1in}{.1in}}
\begin{document}
\preprint{ANL-HEP-PR-11-34 and UMTG-24}
\title[ASFE]{Approximate Solutions of Functional Equations}
\author{Thomas Curtright}
\affiliation{Department of Physics, University of Miami, Coral Gables, FL 33124-8046, USA}
\author{Xiang Jin}
\affiliation{Department of Physics, University of Miami, Coral Gables, FL 33124-8046, USA}
\author{Cosmas Zachos}
\affiliation{High Energy Physics Division, Argonne National Laboratory, Argonne, IL
60439-4815, USA\medskip\medskip}

\begin{abstract}
\medskip

Approximate solutions to functional evolution equations are constructed
through a combination of series and conjugation methods, and relative errors
are estimated. \ The methods are illustrated, both analytically and
numerically, by construction of approximate continuous functional iterates for
$\frac{x}{1-x}$, $\sin x$, and $\lambda x\left(  1-x\right)  $. \ Simple
functional conjugation by these functions, and their inverses, substantially
improves the numerical accuracy of formal series approximations for their
continuous iterates.

\vspace{1in}%
\begin{center}
\includegraphics[
height=3.1073in,
width=4.6561in
]%
{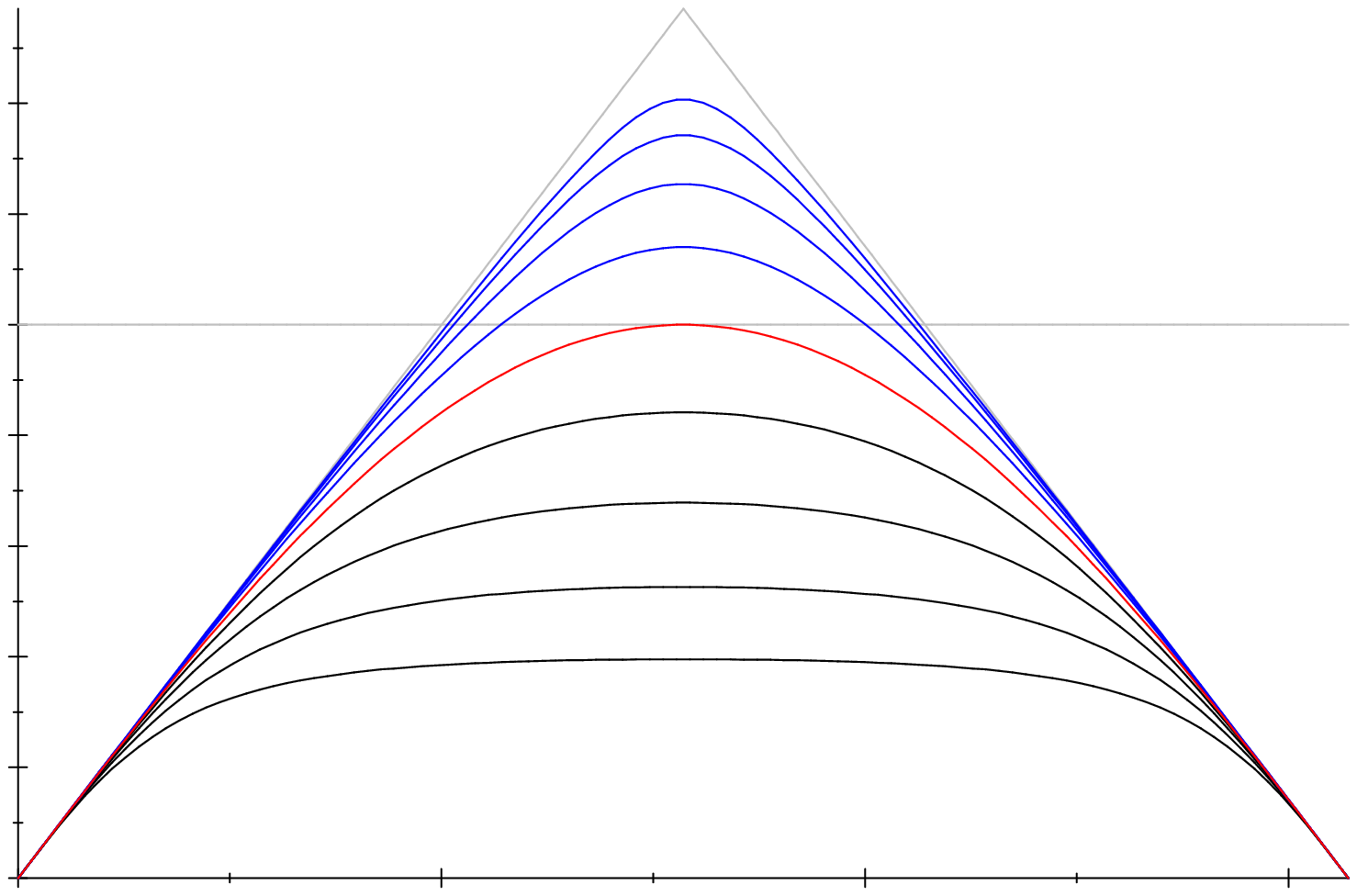}%
\\
Functional iterates of the sine
\end{center}

\end{abstract}
\maketitle

\section{Overview}

There are several circumstances where one encounters the functional evolution
equation \cite{Ac,K},
\begin{equation}
x_{s+t}=x_{s}\circ x_{t}\ , \label{AbelianFE}%
\end{equation}
corresponding to an abelian flow in the subscript parameter, with initial
condition $x_{t=0}=\operatorname*{id}$, the identity map. \ Here,
\textquotedblleft$\circ$\textquotedblright\ is functional composition. \ For
example, such an equation governs the changes in scale that underlie the
renormalization group \cite{G,cz3}, or the rectification of local trajectories
in dynamical systems \cite{Ar}. \ While (\ref{AbelianFE}) does give rise to a
local differential equation, in principle, the actual problem of interest may
instead present the result of a fixed \textquotedblleft
large\textquotedblright\ step in $t$, say $x_{1}$, from which all continuous
iterates $x_{t}$ are to be determined. \ 

In this situation, a useful direct approach is to construct an $N$th order
formal series approximation for $x_{t}$ around a fixed point of $x_{1}$, say
at $x=0$, by series analysis of the $s=1$ case of (\ref{AbelianFE}), written
as%
\begin{equation}
x_{1}\left(  x_{t}\left(  x\right)  \right)  =x_{t}\left(  x_{1}\left(
x\right)  \right)  \ . \label{UnitStepFE}%
\end{equation}
With initial conditions corresponding to $x_{t=0}=\operatorname*{id}$, the
result has the form
\begin{equation}
x_{t}^{\text{(}N\text{ approx)}}\left(  x\right)  =\sum_{k=1}^{N}c_{k}\left(
t\right)  \ x^{k}\ , \label{NSeries}%
\end{equation}
where $c_{1}\left(  0\right)  =1$, $c_{k>1}\left(  0\right)  =0$, so
$x_{t=0}^{\text{(}N\text{ approx)}}\left(  x\right)  =x$. \ Note that it is
only necessary to construct accurate approximations to $x_{t}$ for any unit
interval in $t$, for then the composition rules $x_{1+t}=x_{1}\circ x_{t}$ and
$x_{-1+t}=x_{-1}\circ x_{t}$ can be used to reach higher or lower values of
$t$.

Now, for a given $N$ the series (\ref{NSeries}) may not produce accurate
results for values of $t$, or for an interval of initial $x$, of interest.
\ Quite possibly this may be overcome if the series is convergent, by taking
larger $N$, but if the series is only a formal one, perhaps asymptotic, with
zero radius of convergence, larger $N$ may not be a viable option. \ So what
can be done then?

Formally, for any fixed $N$, with $x_{n}=\underbrace{x_{1}\circ\cdots\circ
x_{1}}_{n\text{ compositions}}$, it follows that%
\begin{equation}
x_{t}=\lim_{n\rightarrow\pm\infty}x_{n}\circ x_{t}^{\text{(}N\text{ approx)}%
}\circ x_{-n}%
\end{equation}
is a solution to (\ref{UnitStepFE}). \ This is a consequence of
\begin{equation}
x_{1}\circ\left(  x_{n}\circ x_{t}^{\text{(}N\text{ approx)}}\circ
x_{-n}\right)  =\left(  x_{n+1}\circ x_{t}^{\text{(}N\text{ approx)}}\circ
x_{-n-1}\right)  \circ x_{1}\ .
\end{equation}
In the limit $n\rightarrow\pm\infty$ Eqn (\ref{UnitStepFE}) is obtained,
\emph{if} either of those limits exists. \ For a specified class of problems
it might be possible to give an elegant proof that either limit exists by
using various fixed point theorems from functional analysis \cite{Ag,Gr}, but
that is \emph{not} our objective here. \ 

The objective here is to estimate the numerical accuracy obtained by
conjugating (\ref{NSeries}) a finite number of times, $n$, with the given,
exactly known, \textquotedblleft finite step\textquotedblright\ functions
$x_{\pm1}$. \ That is, our concern here is the relative error obtained as a
function of $n$, prior to taking the limit $n\rightarrow\infty$. \ In many
cases, $n$-fold conjugation with the given $x_{1}$, and its inverse $x_{-1}$,
\emph{dramatically} improves the numerical accuracy of the series
approximation, with the error vanishing \emph{exponentially} in either $N\ln
n$ or $Nn$. \ This behavior came to light in follow-up studies of earlier work
\cite{cz1,cz2,cv,cz3,c}.

\section{A rational illustration}

As a tractable example \cite{S,J,G,Sz,E}, consider%
\begin{equation}
x_{1}\left(  x\right)  =\frac{x}{1-x}\ ,\ \ \ x_{-1}\left(  x\right)
=\frac{x}{1+x}\ . \label{x1Example}%
\end{equation}
In this particular simple case an exact solution to (\ref{AbelianFE}) is
\begin{equation}
x_{t}\left(  x\right)  =\frac{x}{1-xt}\ . \label{ExactExample}%
\end{equation}
But to illustrate the series method, consider (\ref{UnitStepFE}) given only
$x_{1}$ in (\ref{x1Example}). \ Recursive analysis in powers of $x$
immediately gives $c_{k}\left(  t\right)  =t^{k-1}$, where we have
$\emph{defined}$ the form and scale of the $t~$parameterization by the
\emph{choice} $c_{2}\left(  t\right)  =t$, to be in agreement with
(\ref{ExactExample}). \ For instance, with $x_{t}\left(  x\right)
=x+tx^{2}+c_{3}\left(  t\right)  x^{3}+O\left(  x^{4}\right)  $ we find%
\begin{equation}
\frac{x_{t}\left(  x\right)  }{1-x_{t}\left(  x\right)  }-x_{t}\left(
\frac{x}{1-x}\right)  =\left(  t^{2}-c_{3}\left(  t\right)  \right)
x^{4}+O\left(  x^{5}\right)  \ .
\end{equation}
Thus (\ref{UnitStepFE}) is satisfied if and only if $c_{3}\left(  t\right)
=t^{2}$. \ So it goes to higher orders in $x$, with each $c_{k+1}$ determined
by $c_{j\leq k}$ to satisfy (\ref{UnitStepFE}) --- not just by direct
expansion of (\ref{ExactExample}).

The result for the approximation is therefore%
\begin{equation}
x_{t}^{\text{(}N\text{ approx)}}\left(  x\right)  =\sum_{k=1}^{N}%
t^{k-1}\ x^{k}=\frac{\left(  1-t^{N}x^{N}\right)  x}{1-tx}\ .
\label{RationalNSeries}%
\end{equation}
Then, by $n$-fold conjugation of this approximation with $x_{1}$ and its
inverse, obtain%
\begin{equation}
x_{n}\circ x_{t}^{\text{(}N\text{ approx)}}\circ x_{-n}\left(  x\right)
=\left.  x\left(  1-\left(  \frac{tx}{1+nx}\right)  ^{N}\right)  \right/
\left(  1-tx+nx\left(  \frac{tx}{1+nx}\right)  ^{N}\right)  \ .
\end{equation}
This indeed converges to the exact result, (\ref{ExactExample}), as
$n\rightarrow\infty$ for any \emph{fixed} $N>1$. \ But what is the
\emph{relative error} for finite $n$?

Since we know the exact answer in this elementary case, this\ error is not
difficult to compute. \ For any fixed $N$ and finite $n$ the relative error
is
\begin{equation}
R_{t}\left(  x,N,n\right)  \overset{\text{defn.}}{=}\frac{x_{t}\left(
x\right)  -x_{n}\circ x_{t}^{\text{(}N\text{ approx)}}\circ x_{-n}\left(
x\right)  }{x_{t}\left(  x\right)  }=\frac{1-tx+nx}{1-tx+nx\left(  \frac
{tx}{1+nx}\right)  ^{N}}\left(  \frac{tx}{1+nx}\right)  ^{N}\text{ .}
\label{ExactErrorExample}%
\end{equation}
For $N>1$ this indeed vanishes as $n\rightarrow\infty$, for any fixed $x$, so
long as $tx\neq1$. \ However, more importantly, for $n$ finite but large
compared to both $t$ and $1/x$, this error goes to zero as the $\left(
N-1\right)  $st power of $n$,
\begin{equation}
R_{t}\left(  x,N,n\right)  ~\underset{n\gg\left\vert t\right\vert ,\left\vert
1/x\right\vert }{\approx}~\frac{t^{N}x}{1-tx}~\frac{1}{n^{N-1}}\ .
\label{ErrorExample}%
\end{equation}
Therefore, in principle, one can obtain numerical results to any desired
accuracy by repeated conjugation of the approximate series with the given
step-function $x_{1}$. \ It is remarkable that it is only required to take any
fixed $N\geq2$ to produce such results, although in practice, as is manifest
in (\ref{ErrorExample}), computational efficiencies can be improved by taking
larger $N$ since the desired accuracy is then reached for smaller $n$.

The result for the relative error can be understood in terms of Schr\"{o}der
theory \cite{S}\ as it applies to this simple example. \ The continuous
iterate $x_{t}$ given in (\ref{ExactExample}) may be constructed from a
Schr\"{o}der function, $\Psi$, and its inverse. \ An exact result for this
particular example is given by \cite{Eigenvalue}%
\begin{equation}
\Psi\left(  x\right)  =\exp\left(  -1/x\right)  \ ,\ \ \ \Psi^{-1}\left(
x\right)  =-1/\ln\left(  x\right)  \ , \label{RationalPsi}%
\end{equation}
In Schr\"{o}der's conjugacy framework, the general iterate is $x_{t}\left(
x\right)  =\Psi^{-1}\left(  e^{t}\Psi\left(  x\right)  \right)  $, or,%
\begin{equation}
\Psi\left(  x_{t}\left(  x\right)  \right)  =e^{t}\Psi\left(  x\right)  \ .
\label{RationalPsiEqn}%
\end{equation}
Thus $w\equiv\Psi\left(  x\right)  $ is just the change of variable that
reduces the effect of evolution in $t$ to nothing but a multiplicative
rescaling, $w_{t}=e^{t}w$.

But suppose that the exact $x_{t}$ were supplanted by an approximation of the
form $x_{t}^{\text{(}N\text{ approx)}}=x_{t}+\alpha x_{t}^{N+1}+O\left(
x_{t}^{N+2}\right)  $, for some coefficient $\alpha$. \ Then
\begin{equation}
\Psi\left(  x_{t}^{\text{(}N\text{ approx)}}\right)  =\left(  1+\alpha
x_{t}^{N-1}+O\left(  x_{t}^{N}\right)  \right)  \exp\left(  -1/x_{t}\right)
\ .
\end{equation}
Alternatively, with $x_{t}=-1/\ln\Psi\left(  x_{t}\right)  =-1/\ln w_{t}$,
\begin{equation}
w_{t}^{\text{(approx)}}\equiv\Psi\left(  x_{t}^{\text{(}N\text{ approx)}%
}\left[  w_{t}\right]  \right)  =\left(  1+\frac{\alpha}{\left(  -\ln
w_{t}\right)  ^{N-1}}+O\left(  \frac{1}{\left(  -\ln w_{t}\right)  ^{N}%
}\right)  \right)  w_{t}\ .
\end{equation}
Then it follows from the multiplicative rescaling behavior of $w$ that under
the variable change $x_{t}\rightarrow w_{t}$ the conjugated approximation
$x_{n}\circ x_{t}^{\text{approx}}\circ x_{-n}$ presents itself as
\begin{equation}
w_{n}\circ w_{t}^{\text{approx}}\circ w_{-n}=\left(  1+\frac{\alpha}{\left(
n-\ln w_{t}\right)  ^{N-1}}+O\left(  \frac{1}{\left(  n-\ln w_{t}\right)
^{N}}\right)  \right)  w_{t}\ .
\end{equation}
This gives a relative error with the same power law asymptotics \cite{F1} as
(\ref{ErrorExample}), namely, $R\underset{n\rightarrow\infty}{\sim}%
\alpha/n^{N-1}$.

\section{The roots of sin}

A more interesting example is provided by the sine function, where our
notation for the continuous iterate is now $\sin_{t}\left(  x\right)  $ with
$\sin_{1}\left(  x\right)  =\sin\left(  x\right)  $ and $\sin_{-1}\left(
x\right)  =\arcsin\left(  x\right)  $. \ In this notation, the abelian
functional composition equation is%
\begin{equation}
\sin_{s}\circ\sin_{t}=\sin_{s+t}\ . \label{AbelianSine}%
\end{equation}
Again specializing to $s=1$, written as%
\begin{equation}
\sin\left(  \sin_{t}\left(  x\right)  \right)  =\sin_{t}\left(  \sin\left(
x\right)  \right)  \ , \label{UnitStepSine}%
\end{equation}
we find a formal series solution,%
\begin{equation}
\sin_{t}^{\text{(approx)}}\left(  x\right)  =x-\frac{1}{3!}tx^{3}%
+\frac{\left(  5t-4\right)  t}{5!}x^{5}-\frac{\left(  175t^{2}%
-336t+164\right)  t}{3\times7!}x^{7}+\frac{\left(  25t-24\right)  \left(
8-7t\right)  ^{2}t}{9!}x^{9}+O\left(  x^{11}\right)  \ . \label{SineSeries}%
\end{equation}
This is sufficient for the numerical work to follow\ (other results for this
example are available online:
\ \href{http://server.physics.miami.edu/~curtright/Schroeder.html}{Schroeder.html}
\&
\href{http://server.physics.miami.edu/~curtright/TheRootsOfSin.pdf}{TheRootsOfSin.pdf}%
).

In this case, the series (\ref{SineSeries}) is probably \emph{not} convergent
for almost all $t$ \cite{Si,Sz,E}, although it obviously is convergent for
$t\in%
\mathbb{Z}
$. \ Rather, for generic $t$ the series appears to be asymptotic. \ For
example, for $t=1/2$, using Mathematica to extend the series (\ref{SineSeries}%
) to $O\left(  x^{81}\right)  $ or so, the smooth behavior of the series
coefficients $c_{k}$ for $k\leq61$ suggests a finite radius of convergence
$\approx4/3$, as estimated by $1/\left\vert c_{k}\right\vert ^{1/k}$. \ But
then less smooth behavior sets in for the $c_{k>61}$ and numerical estimates
of the radius begin to fall towards zero, as would be expected for an
asymptotic series. \ 

Also note in passing that (\ref{SineSeries}) implies the corresponding
Schr\"{o}der function has an essential singularity at $x=0$, as given
explicitly by \cite{Eigenvalue}
\begin{equation}
\Psi\left(  x\right)  \underset{x\rightarrow0}{\sim}x^{6/5}\exp\left(
\frac{3}{x^{2}}+\frac{79}{1050}x^{2}+\frac{29}{2625}x^{4}+O\left(
x^{5}\right)  \right)  \ , \label{SinePsi}%
\end{equation}
to be compared to (\ref{RationalPsi}). \ This follows from $\Psi\left(
x\right)  =\exp\left(  \int dx/v\left(  x\right)  \right)  $, using $v\left(
x\right)  =\left.  d\sin_{t}\left(  x\right)  /dt\right\vert _{t=0}%
\approx\left.  d\sin_{t}^{\text{(approx)}}\left(  x\right)  /dt\right\vert
_{t=0}$.

Nonetheless, the conjugation method produces approximations $\sin_{t}\approx
A_{n,t}$ which provide compelling numerical evidence for the existence of a
limit, hence an exact result for $\sin_{t}$, as $n\rightarrow\infty$. \ But
note it is important in this case to take the conjugations to be of the form
\begin{equation}
A_{n,t}=\sin_{-n}\circ\sin_{t}^{\text{(approx)}}\circ\sin_{n}\ ,
\label{SineApproxs}%
\end{equation}
the general rule being to act first with functions that are \emph{smaller} in
magnitude than the identity map, thereby leading to evaluation of the
truncated series at points closer to $x=0$. \ 

The improvement wrought by conjugation of the truncated ninth-order series in
(\ref{SineSeries}) is easily seen in the following graphs, for the case
$t=1/2$. \ The series itself is not credible beyond $x=\pi/2$, but a single
conjugation forces the expected periodicity in $x$, and gives the more
plausible green curve seen in the first Figure. \ Repeated conjugation does
not produce any discernible differences with the green curve to the accuracy
of that first Figure, but when the graph is magnified, as it is in the second
Figure, numerical evidence for convergence of the sequence of conjugations is
quite compelling.

\noindent%
{\parbox[b]{3.2879in}{\begin{center}
\includegraphics[
height=2.0097in,
width=3.2879in
]%
{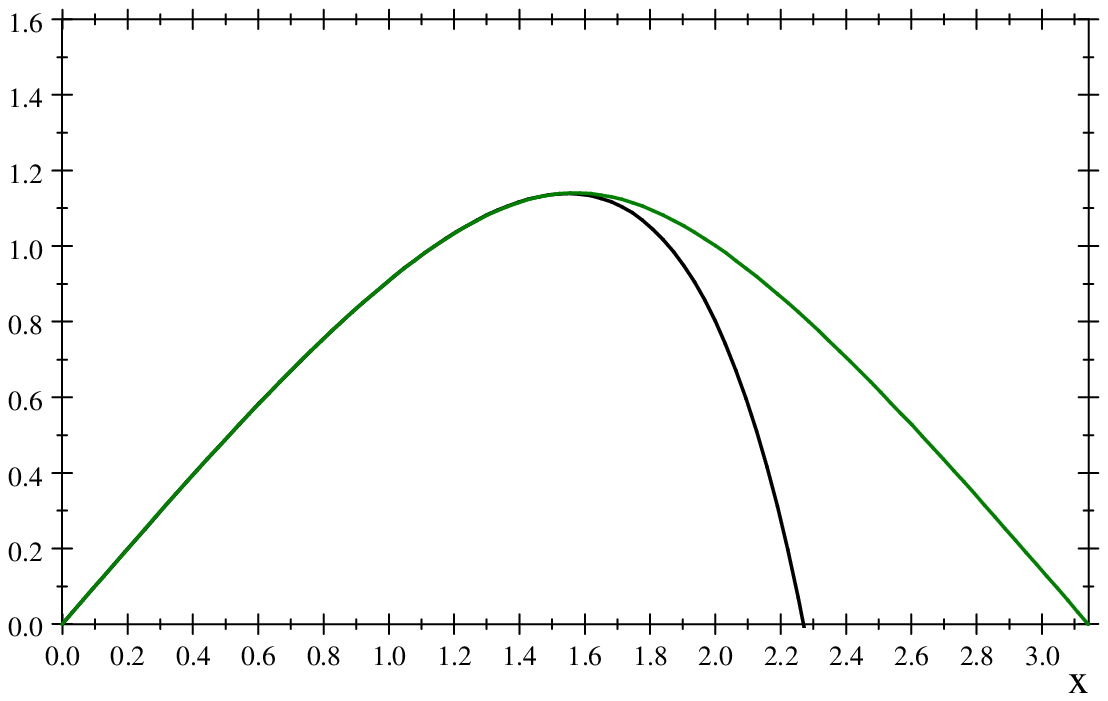}%
\\
The ninth order series (\ref{SineSeries}) for $t=1/2$, in black, along with
its $n=1$ conjugation (\ref{SineApproxs}), in green.
\end{center}}}%
\
{\parbox[b]{3.2879in}{\begin{center}
\includegraphics[
height=2.0088in,
width=3.2879in
]%
{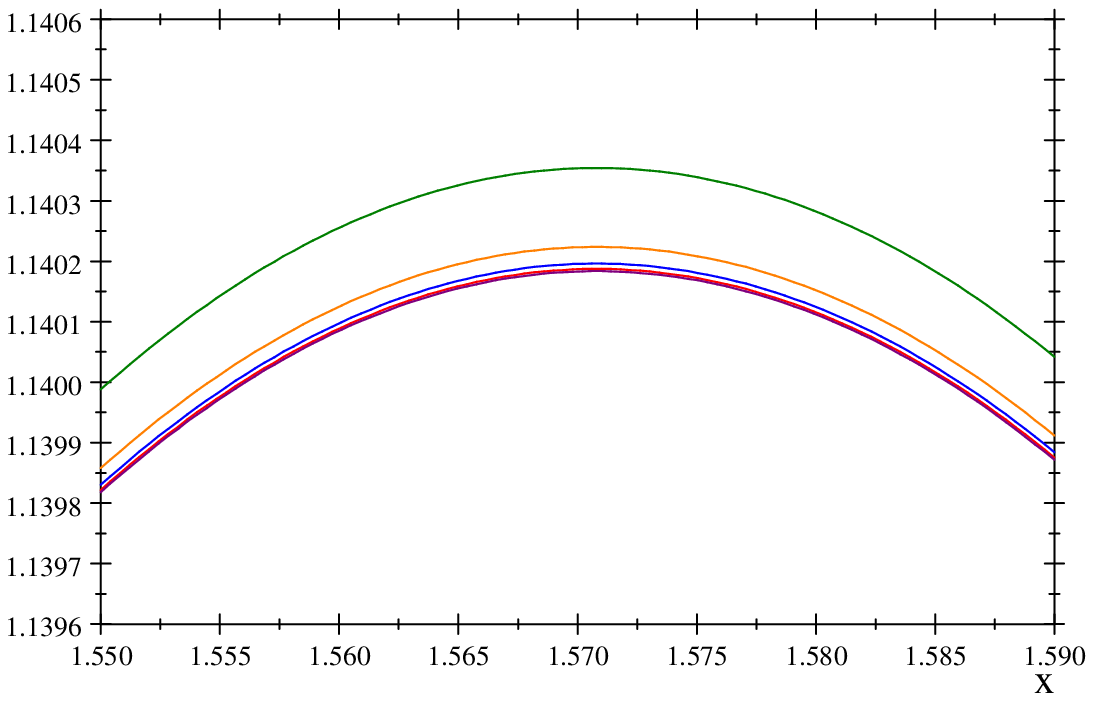}%
\\
The $t=1/2$, $n=1,2,3,4,$ \& $5$-fold conjugations (\ref{SineApproxs}), in
green, orange, blue, red, and purple, respectively.
\end{center}}}%

Proceeding in this way leads to the set of curves for various values of $t$
shown in the following Figure. \ Each of these curves results from the
five-fold conjugation $A_{5,t}$ of the truncated ninth-order series
(\ref{SineSeries}). \ Note once again, as previously remarked in a general
context, it is only necessary to construct accurate approximations to
$\sin_{t}$ for any unit interval in $t$, for then the composition rules
$\sin_{1+t}=\sin\circ\sin_{t}$ and $\sin_{-1+t}=\arcsin\circ\sin_{t}$ can be
used to construct the curves for higher or lower values of $t$. \ Also note
from the numerics the obvious inference that $\sin_{t}\left(  x\right)  $
becomes the periodic triangular \textquotedblleft sawtooth\textquotedblright%
\ function as $t\rightarrow0$, with $\lim_{t\rightarrow0}\sin_{t}\left(
\left(  2k+1\right)  \pi/2\right)  =\left(  -1\right)  ^{k}\pi/2$.%
\begin{center}
\includegraphics[
height=3.9172in,
width=6.2744in
]%
{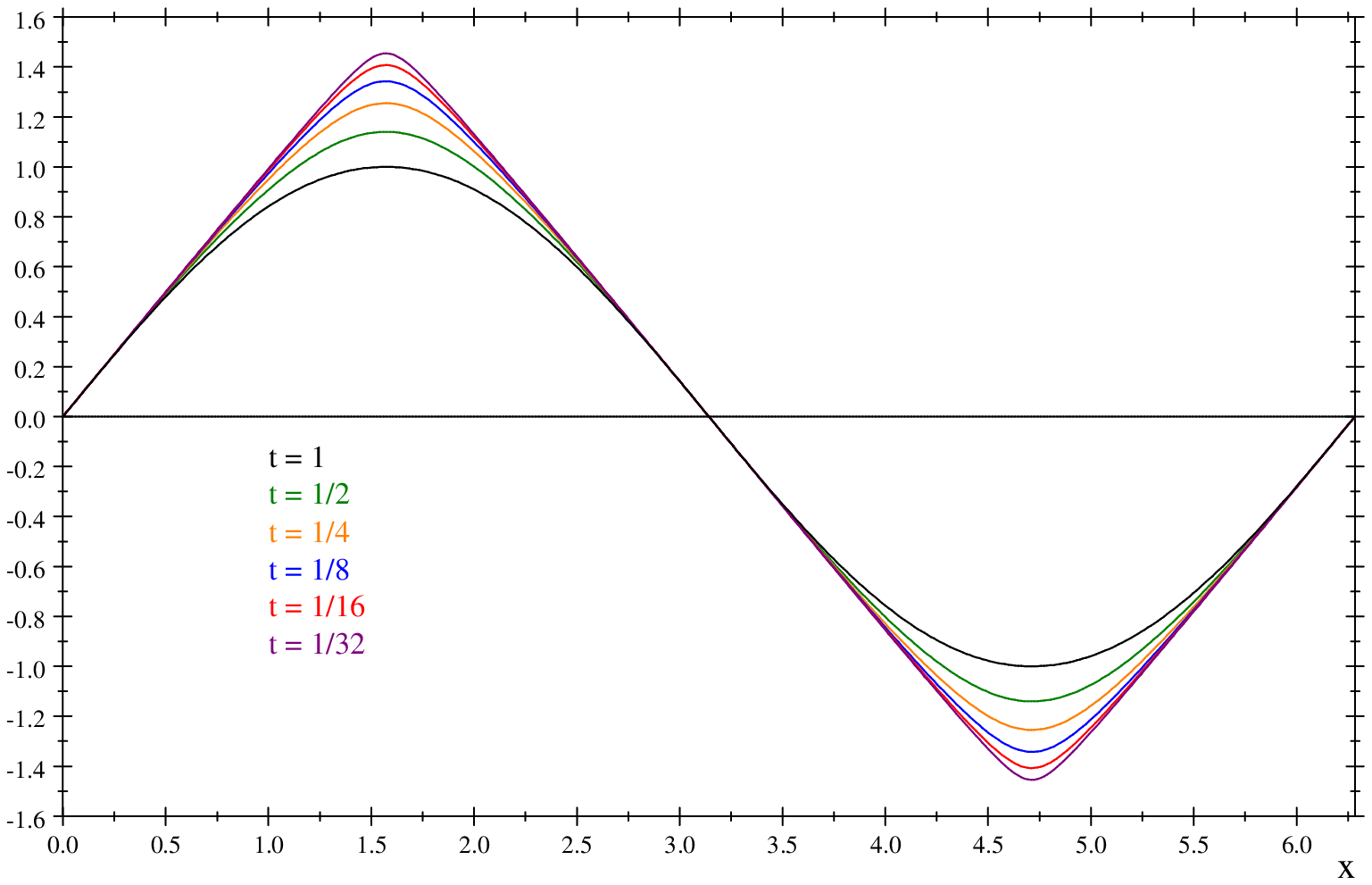}%
\\
Various $\sin_{t}\left(  x\right)  $ as given by five-fold conjugations
$A_{5,t}\left(  x\right)  $ of the ninth-order series (\ref{SineSeries}), for
$t\leq1$.
\end{center}

As constructed, $\sin_{t}\left(  x\right)  $ for $t\geq0$ is guaranteed to be
real for all real values of $x$, but certainly it is not obvious for generic
$t$ what numerical values are actually attained at the extrema for $x=\frac
{1}{2}\pi\operatorname{mod}\pi$. \ It suffices here for us to point out that
the maxima are approximated by the simple expression:%
\begin{equation}
\sin_{t}\left(  \pi/2\right)  \approx\left(  \frac{\pi}{2}\right)
^{1-\sqrt{t}}\ . \label{ApproxMax}%
\end{equation}
At least, this is true for $0\leq t\leq1$, where the relative error between
the exact (numerical) value of $\sin_{t}\left(  \pi/2\right)  $ and this
approximation is less than about 3 parts per mille. \ (More accurate numerics
are available
\href{http://server.physics.miami.edu/curtright/SineMapIterationExtrema.pdf}{online}%
.) \ The branch point at $t=0$ exhibited in this approximate expression is
perhaps the most direct numerical evidence that the iterates are \emph{not}
analytic at $t=0$ for all $x$.

The graphs in the second Figure above give a sense of the overall relative
error, but lacking closed-form expressions for either the iterates or the
conjugations of the series approximations, closed-form results for the error
are not available for generic $t$. \ For $t\in%
\mathbb{Z}
$, however, precise calculation is indeed possible since both exact results
and convergent series are known. \ It suffices here to consider just one exact
case, $t=1$. \ Defining the relative error as before,%
\begin{equation}
R_{1}\left(  x,n\right)  =\frac{\sin x-\sin_{-n}\circ\sin_{1}^{\text{(approx)}%
}\circ\sin_{n}\left(  x\right)  }{\sin x}\ ,
\end{equation}
we have computed numerically the error involved in various conjugations of the
ninth-order series. \ As previously remarked, conjugation by the sine
guarantees the approximations are periodic. \ The results are shown here.%
\begin{center}
\includegraphics[
height=1.6398in,
width=4.5258in
]%
{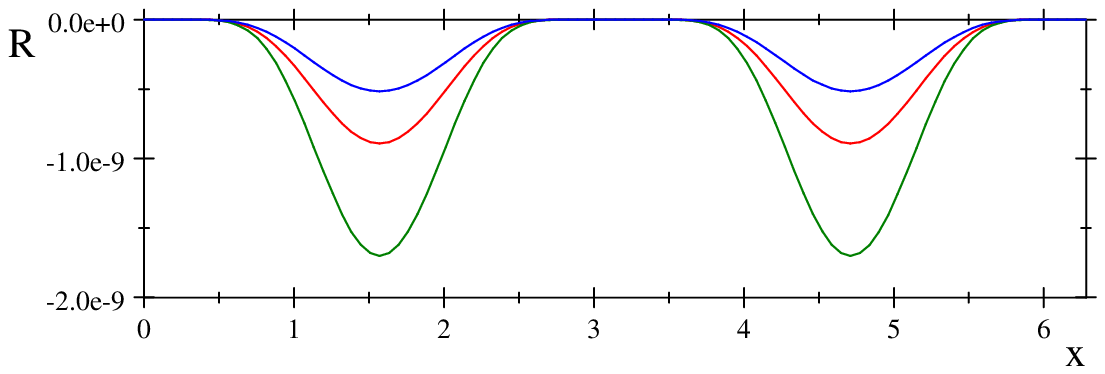}%
\\
$R_{t=1}\left(  x,n\right)  $ for $n=4,$ $5,$ and $6$, in green, red, and
blue, respectively.
\end{center}

It would be interesting to compute relative errors for other, generic $t$, but
at this stage it is only possible for us to compute the \emph{relative
successive differences},%
\begin{equation}
S_{t}\left(  x,n\right)  =\frac{\sin_{-n}\circ\sin_{t}^{\text{(approx)}}%
\circ\sin_{n}\left(  x\right)  -\sin_{-n+1}\circ\sin_{t}^{\text{(approx)}%
}\circ\sin_{n-1}\left(  x\right)  }{\sin_{-n}\circ\sin_{t}^{\text{(approx)}%
}\circ\sin_{n}\left(  x\right)  }\ .
\end{equation}
For example, consider $t=1/2$ using the ninth-order series.%
\begin{center}
\includegraphics[
height=1.6398in,
width=4.5258in
]%
{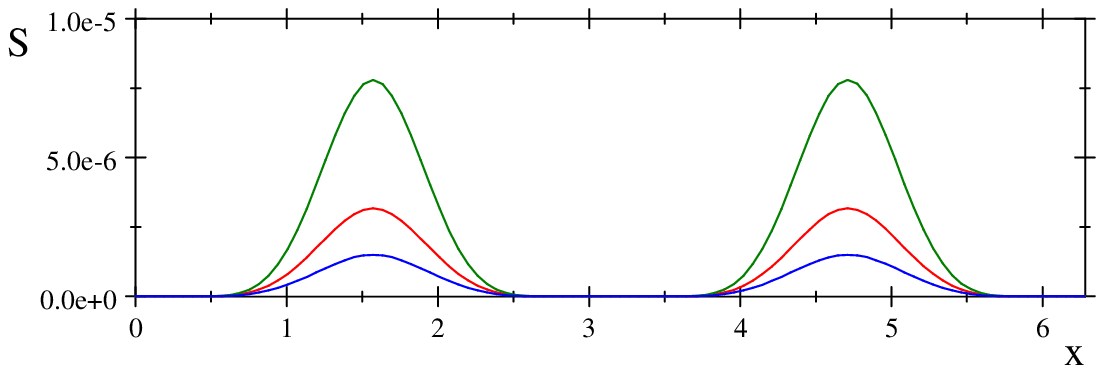}%
\\
$S_{t=1/2}\left(  x,n\right)  $ for $n=4,$ $5,$ and $6$, in green, red, and
blue, respectively.
\end{center}
This numerical data supports the proposition that these continuous iterates of
the sine function are well-defined and straightforward to compute using the
methods in this paper.

\section{The logistic map}

The sequence of conjugations converges more rapidly in situations where the
underlying Schr\"{o}der function is analytic about the fixed point, in
contrast to (\ref{RationalPsi}) and (\ref{SinePsi}). \ Instead of power law
behavior, for such situations the relative error vanishes exponentially in $n
$, the number of conjugations. \ The general theory is well-illustrated by the
logistic map,%
\begin{equation}
x_{1}\left(  x\right)  =\lambda x\left(  1-x\right)  \ ,\ \ \ \text{for
\ \ }0\leq x\leq\lambda/4\text{ \ \ and \ }0\leq\lambda\leq4\text{.}%
\end{equation}
The result of the theory is the following:

\noindent\textbf{[Theorem]} \ Relative error after $n$-fold conjugation of the
truncated series (\ref{NSeries}) is given by%
\begin{align}
R_{t}\left(  x,\lambda,N,n\right)   &  =\frac{x_{t}\left(  x\right)
-x_{n}\left(  x_{t}^{\text{(}N\text{ approx)}}\left(  x_{-n}\left(  x\right)
\right)  \right)  }{x_{t}\left(  x\right)  }\nonumber\\
&  =\left(  \frac{1}{\lambda}\right)  ^{nN}r\left(  x,t,\lambda,N,n\right)
\nonumber\\
&  =\left(  \frac{1}{\lambda}\right)  ^{nN}\lambda^{-t}c_{N+1}\left(
t,\lambda\right)  ~x^{N}+O\left(  x^{N+1}\right)  \ .
\end{align}
\noindent\textbf{[Proof]} \ To sketch a proof, and to see more clearly the
assumptions involved, as well as to obtain expressions for $r\left(
x,t,\lambda,N,n\right)  $, write the truncated series as%
\begin{equation}
x_{t}^{\text{(}N\text{ apx)}}\left(  x\right)  =x_{t}\left(  x\right)
-x^{N+1}\delta_{N}\left(  x,t\right)  \ , \label{DeltaDefn}%
\end{equation}
where $\delta_{N}\left(  x,t\right)  $ represents the exact difference, whose
expansion in $x$ begins $O\left(  x^{0}\right)  $. \ Thus the conjugation
gives exactly%
\begin{equation}
x_{n}\left(  x_{t}^{\text{(}N\text{ apx)}}\left(  x_{-n}\left(  x\right)
\right)  \right)  =x_{n}\left(  x_{t}\left(  x_{-n}\left(  x\right)  \right)
-\left(  x_{-n}\left(  x\right)  \right)  ^{N+1}\delta_{N}\left(
x_{-n}\left(  x\right)  ,t\right)  \right)  \ .
\end{equation}
Now expand the RHS in powers of $\left(  x_{-n}\right)  ^{N+1}\delta_{N}$,%
\begin{align}
x_{n}\left(  x_{t}\left(  x_{-n}\left(  x\right)  \right)  -\left(
x_{-n}\left(  x\right)  \right)  ^{N+1}\delta_{N}\left(  x_{-n}\left(
x\right)  ,t\right)  \right)   &  =x_{n}\left(  x_{t}\left(  x_{-n}\left(
x\right)  \right)  \right) \\
&  -\left(  x_{-n}\left(  x\right)  \right)  ^{N+1}~\delta_{N}\left(
x_{-n}\left(  x\right)  ,t\right)  ~x_{n}^{\prime}\left(  x_{t}\left(
x_{-n}\left(  x\right)  \right)  \right)  +O\left(  \left(  x_{-n}^{2}\right)
^{N+1}\delta_{N}^{2}\right)  \ .\nonumber
\end{align}
Since it consists of exact trajectories that obey (\ref{AbelianFE}), the first
term gives $x_{n}\left(  x_{t}\left(  x_{-n}\left(  x\right)  \right)
\right)  =x_{t}\left(  x\right)  $, while the second term involves%
\begin{equation}
x_{n}^{\prime}\left(  x_{t}\left(  x_{-n}\left(  x\right)  \right)  \right)
=\dfrac{1}{\dfrac{d}{dx}x_{t}\left(  x_{-n}\left(  x\right)  \right)  }%
\frac{d}{dx}x_{n}\left(  x_{t}\left(  x_{-n}\left(  x\right)  \right)
\right)  =\frac{1}{\dfrac{d}{dx}x_{t-n}\left(  x\right)  }\frac{d}{dx}%
x_{t}\left(  x\right)  \ ,
\end{equation}
again using (\ref{AbelianFE}). \ Thus%
\begin{equation}
x_{t}\left(  x\right)  -x_{t}^{\text{(}N\text{ apx }\left[  n\right]
\text{)}}\left(  x\right)  =\left(  x_{-n}\left(  x\right)  \right)
^{N+1}~\delta_{N}\left(  x_{-n}\left(  x\right)  ,t\right)  ~\frac
{dx_{t}\left(  x\right)  /dx}{dx_{t-n}\left(  x\right)  /dx}+O\left(  \left(
x_{-n}^{2}\right)  ^{N+1}\delta_{N}^{2}\right)  \ .
\end{equation}

To proceed, we require that the unit step function is such that $x_{-n}\left(
x\right)  $ flows toward the fixed point at the origin for the point $x$ under
consideration (so we suppose $\left\vert \lambda\right\vert >1$, but if not,
just interchange $x_{1}$ and $x_{-1}$). \ We also suppose that $n$ has been
chosen large compared to $t$ so that
\begin{equation}
x_{t-n}\left(  x\right)  \equiv\lambda^{t-n}\varepsilon_{n}\left(  x,t\right)
\label{EpsilonDefn}%
\end{equation}
is \emph{small}, where $\varepsilon_{n}\left(  x,t\right)  =x+O\left(
x^{2}\right)  $. \ If these requirements are met, then%
\begin{align}
x_{t}\left(  x\right)  -x_{t}^{\text{(}N\text{ apx }\left[  n\right]
\text{)}}\left(  x\right)   &  =\lambda^{-n\left(  N+1\right)  }%
\varepsilon_{n}^{N+1}\left(  x,0\right)  ~\delta_{N}\left(  \lambda
^{-n}\varepsilon_{n}\left(  x,0\right)  ,t\right)  ~\frac{dx_{t}\left(
x\right)  /dx}{\lambda^{t-n}d\varepsilon_{n}\left(  x,t\right)  /dx}%
\nonumber\\
&  +O\left[  \lambda^{-2n\left(  N+1\right)  }\varepsilon_{n}^{2N+2}\delta
_{N}^{2}\right]  \ , \label{Difference}%
\end{align}
where $d\varepsilon_{n}\left(  x,t\right)  /dx=1+O\left(  x\right)  $. \ The
result for the relative error is therefore of the form in the statement of the
Theorem, with
\begin{equation}
r\left(  x,t,\lambda,N,n\right)  =\frac{\varepsilon_{n}^{N+1}\left(
x,0\right)  ~\delta_{N}\left(  \lambda^{-n}\varepsilon_{n}\left(  x,0\right)
,t\right)  }{\lambda^{t}d\varepsilon_{n}\left(  x,t\right)  /dx}~\frac
{dx_{t}\left(  x\right)  /dx}{x_{t}\left(  x\right)  }+O\left[  \lambda
^{-n\left(  N+2\right)  }\varepsilon_{n}^{2N+2}\delta_{N}^{2}\right]  \ .
\label{Remainder}%
\end{equation}
For small $x$ we have%
\begin{align}
\varepsilon_{n}^{N+1}\left(  x,0\right)   &  =x^{N+1}\left(  1+O\left(
x\right)  \right)  \ ,\ \ \ d\varepsilon_{n}\left(  x,t\right)  /dx=1+O\left(
x\right)  \ ,\nonumber\\
\delta_{N}\left(  \lambda^{-n}\varepsilon_{n}\left(  x,0\right)  ,t\right)
&  =c_{N+1}\left(  t,\lambda\right)  +O\left(  x\right)  \ ,\ \ \ \frac
{dx_{t}\left(  x\right)  /dx}{x_{t}\left(  x\right)  }=\frac{1}{x}\left(
1+O\left(  x\right)  \right)  \ ,
\end{align}
and therefore%
\begin{equation}
r\left(  x,t,\lambda,N,n\right)  =\lambda^{-t}c_{N+1}\left(  t,\lambda\right)
~x^{N}+O\left(  x^{N+1}\right)  \ , \label{xSmallRemainder}%
\end{equation}
again as previously stated. $\ \blacksquare$

The form given in (\ref{Remainder}) enables analysis of the size of the error
as a function of $x$. \ Moreover, the form in (\ref{xSmallRemainder})
\emph{suggests} an approximate scaling law for the error%
\begin{equation}
R_{t}\left(  x,\lambda,N,n\right)  \approx\lambda^{N}R_{t}\left(
x,\lambda,N,n+1\right)  \ , \label{ScalingLaw}%
\end{equation}
at least for $x$ near the origin. \ It is interesting to check whether this is
true for larger $x$. \ In fact, it is.

To be specific, consider the case $\lambda=2$ which can be solved in
closed-form to obtain:%
\begin{align}
\Psi\left(  x\right)   &  =-\frac{1}{2}\ln\left(  1-2x\right)  \ ,\ \ \ \Psi
^{-1}\left(  x\right)  =\frac{1}{2}\left(  1-e^{-2x}\right)  \ ,\\
x_{t}\left(  x\right)   &  =\frac{1}{2}\left(  1-\left(  1-2x\right)  ^{2^{t}%
}\right)  \ ,\ \ \ \frac{dx_{t}\left(  x\right)  }{dx}=2^{t}\left(
1-2x\right)  ^{-1+2^{t}}\ ,\label{2Exact}\\
\varepsilon_{n}\left(  x,t\right)   &  =2^{n-t-1}\left(  1-\left(
1-2x\right)  ^{2^{t-n}}\right)  \ ,\ \ \ \frac{d\varepsilon_{n}\left(
x,t\right)  }{dx}=\left(  1-2x\right)  ^{-1+2^{t-n}}\ ,
\end{align}
as well as%
\begin{align}
\varepsilon_{n}^{N+1}\left(  x,0\right)   &  =\left(  2^{n}\frac{1}{2}\left(
1-\left(  1-2x\right)  ^{2^{-n}}\right)  \right)  ^{N+1}\ ,\\
\delta_{N}\left(  x,t\right)   &  =\left(  \frac{\left(  -2\right)  ^{N}%
}{\left(  N+1\right)  !}%
{\displaystyle\prod\limits_{j=0}^{N}}
\left(  2^{t}-j\right)  \right)  \times\operatorname{hypergeom}\left(  \left[
1,N+1-2^{t}\right]  ,\left[  2+N\right]  ,2x\right)  \ .
\end{align}
This last result may be obtained by solving (\ref{UnitStepFE}) recursively for
$x_{1}\left(  x\right)  =2x\left(  1-x\right)  $ --- not just by direct
expansion of (\ref{2Exact}) --- to obtain coefficients%
\begin{equation}
c_{k}\left(  t\right)  =\frac{\left(  -2\right)  ^{k-1}}{k!}%
{\displaystyle\prod\limits_{j=0}^{k-1}}
\left(  2^{t}-j\right)  \ .
\end{equation}

These $\lambda=2$ results are all well-behaved enough for the steps in the
proof of the Theorem to be valid for $x<1/2$. \ However, at the upper end of
the interval, $0\leq x\leq1/2$, some additional consideration is needed. \ The
expression in (\ref{EpsilonDefn}) is $1/2$ at $x=1/2$, independent of $t$ and
$n$, and therefore \emph{not} small. \ That is to say, at $x=1/2$ the
$2^{-nN}$ prefactor in $R_{t}\left(  x,\lambda,N,n\right)  $ is not present to
suppress the error. \ On the other hand, the ratio $\left(  dx_{t}\left(
x\right)  /dx\right)  /\left(  d\varepsilon_{n}\left(  x,t\right)  /dx\right)
$ in (\ref{Difference}), and in (\ref{Remainder}), always vanishes at $x=1/2$,
for any $t$,\ so the leading contribution to the relative error is actually
zero at that point. \ Thus the upper end of the $x$ interval does not pose a
problem after all. \ In fact, as is evident in the graphs to follow, or by
direct calculation, the maximum magnitude of the leading contribution to $R$
occurs for $x<1/2$, for which the $2^{-nN}$ prefactor is present.

Putting all this together for the $\lambda=2$ logistic map, the leading
approximation to the relative error is%
\begin{equation}
R_{t}\left(  x,2,N,n\right)  \approx\delta_{N}\left(  \frac{1}{2}\left(
1-\left(  1-2x\right)  ^{2^{-n}}\right)  ,t\right)  \times\frac{2^{n-N}\left(
1-\left(  1-2x\right)  ^{2^{-n}}\right)  ^{N+1}\times\left(  1-2x\right)
^{2^{t}\left(  1-2^{-n}\right)  }}{\left(  1-\left(  1-2x\right)  ^{2^{t}%
}\right)  }\ . \label{2LeadApprox}%
\end{equation}
For comparison purposes, we plot this last expression for various $N$ and $n$,
and selected $t$, especially to check (\ref{ScalingLaw}), for $\lambda=2.$
\ That scaling law is seen to be hold fairly well, even for $x\approx1/2$. \ 

\noindent%
{\parbox[b]{6.208in}{\begin{center}
\includegraphics[
trim=0.000000in 0.000000in -0.011141in 0.011951in,
height=2.8073in,
width=6.208in
]%
{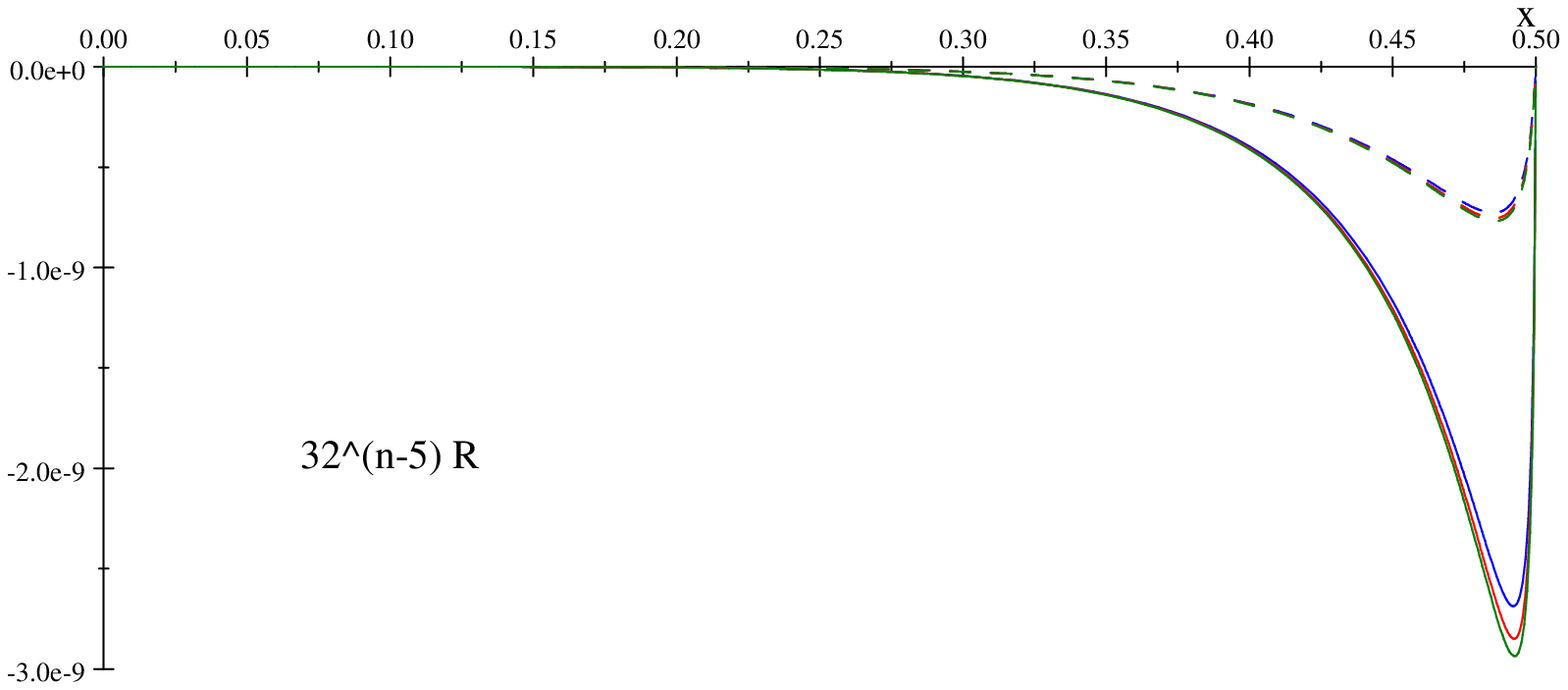}%
\\
Leading approximations to $2^{5\left(  n-5\right)  }R_{t}\left(
x,\lambda=2,N=5,n\right)  $ for $t=1/2$ (solid) and $t=3/4$ (dashed) with
$n=5,\ 6,$ \& $7$, in blue, red, \& green, respectively.
\end{center}}}%

Finally, we note there is no discernible difference, to the accuracy of this
last plot, between the leading approximations and the exact results for the
relative error as computed from (\ref{2Exact}) and the $5$-, $6$-, and
$7$-fold conjugations of $x_{t}^{\text{(}5\text{ approx)}}\left(  x\right)  $.
\ The largest of these differences, between the exact and the leading
approximation of the relative error for $N=5,n=5$, is shown next.

\noindent%
{\parbox[b]{6.2089in}{\begin{center}
\includegraphics[
height=2.8037in,
width=6.2089in
]%
{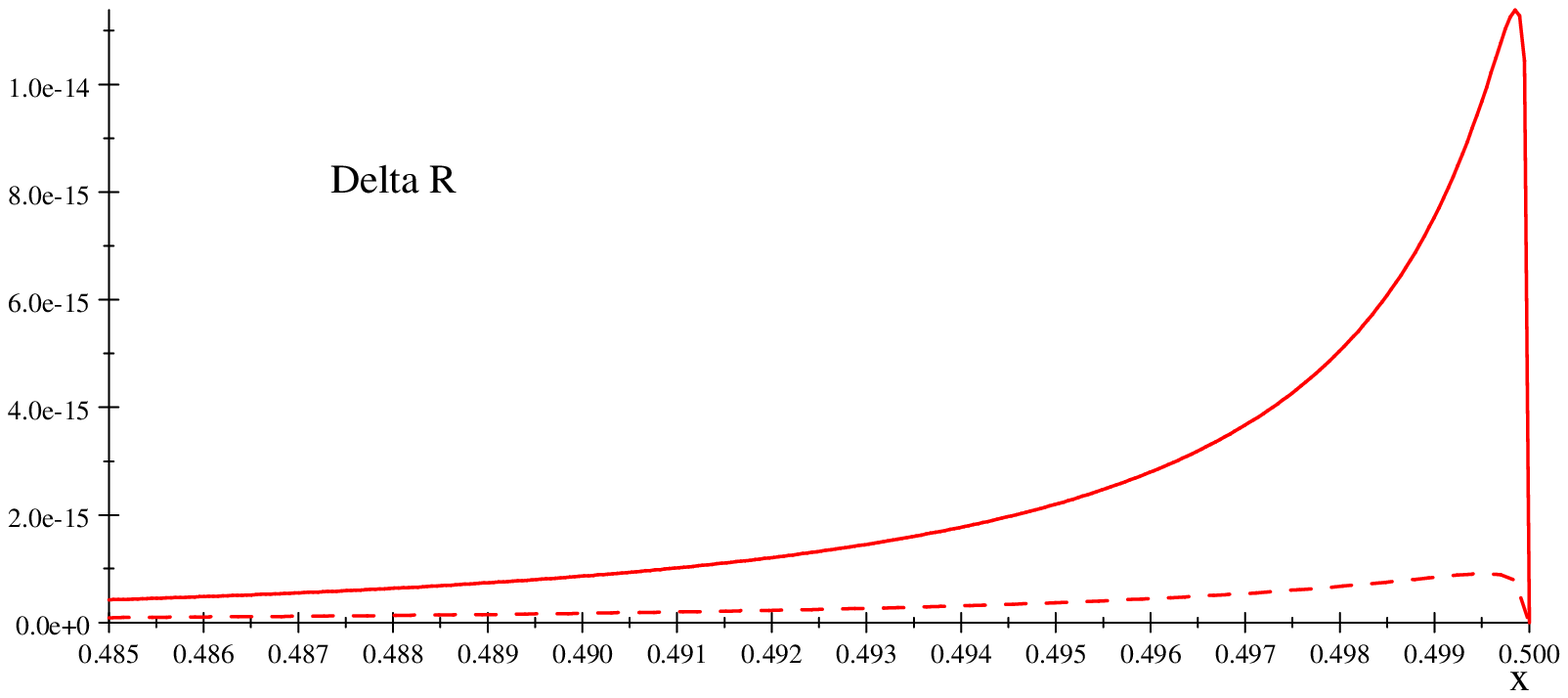}%
\\
$\Delta R=\left.  R_{t}\left(  x,\lambda=2,N=5,n=5\right)  \right\vert
_{\text{exact}}-\left.  R_{t}\left(  x,\lambda=2,N=5,n=5\right)  \right\vert
_{\text{leading approx}}$ for $t=1/2$ (solid) and $t=3/4$ (dashed).
\end{center}}}%

\noindent So, the exact relative errors are essentially indistinguishable from
their leading approximations, at least for $\lambda=2$ and $t=1/2$ \& $t=3/4$,
and the relative error after $7$ conjugations of the $5$th order series is
always less than $3$ parts in $10^{12}$ for these two values of $t$. \ Other
values of $t$ taken from the unit interval $\left[  0,1\right]  $ are
similarly well-approximated by the combined series and conjugation methods. \ 

More or less the same results can be obtained for other values of $\lambda$ as
well, but only for one other case, namely $\lambda=4$, is it possible to
compare to exact, closed-form results. \ 

\section{Summary}

By combining series approximations with functional conjugations, accurate
representations of continuous functional iterates were obtained for selected
rational, sine, and logistic maps, and relative errors were estimated. \ These
examples illustrate both the simplicity and the power of the methods.
\ Although the rational and sine examples have more singular underpinnings
(their Schr\"{o}der functions have essential singularities) than the selected
logistic map (whose Schr\"{o}der function\ is analytic), nevertheless, the
same methods work well for all three cases. \ 

This should be true for many other examples. \ Indeed, it would appear to be a
relatively straightforward task to construct accurate approximations for the
continuous iterations of \emph{all} the classical special functions. \ Many of
these, like the sine iterates depicted above, are expected to exhibit smooth,
intuitive iterations that are easy to grasp, conceptually. \ Others, such as
occur for the logistic map when the parameter leads to chaotic behavior, are
expected to lead to much more exotic iterations. \ Despite the long history of
and voluminous literature on functional equations (see \cite{Ac,K} for history
and bibliographies), the full landscape of features to be encountered is still
largely unexplored, in our opinion. \ Perhaps the methods elucidated in this
paper will help to continue that exploration.

$\mathbf{Acknowledgements}$ \ \textit{We thank David Fairlie and Andrzej
Veitia for discussions related to this research. \ We also thank an anonymous
referee for asking a question which led us to obtain (\ref{ApproxMax}). \ This
work was supported in part by NSF Award 0855386, and in part by the U.S.
Department of Energy, Division of High Energy Physics, under contract
DE-AC02-06CH11357.}


\begin{thebibliography}{99}
\bibitem {Eigenvalue}This result is for a particular choice (namely,
$\lambda=e$) of the \textquotedblleft multiplier\textquotedblright\ in
Schr\"{o}der's equation, $\lambda\Psi\left(  x\right)  =\Psi\left(
x_{1}\left(  x\right)  \right)  $. \ The multiplier is undetermined when
$\left.  d^{n}\Psi\left(  x\right)  /dx^{n}\right\vert _{x=0}$ is zero, or
undefined, for all $n$. \ In such cases it is usually possible to choose
$\lambda$ just as a matter of taste.\smallskip
\end{thebibliography}
\end{document}